\documentclass[12pt]{article}
\usepackage{a41} %
\usepackage{float} %
\usepackage{lscape} %
\usepackage{epsfig} %
\usepackage{amssymb} %
\usepackage{amsmath} %   
\usepackage{verbatim} %
\usepackage{cite} %
\newcommand{\fs}{\footnotesize}

\newcommand{\beq}{\begin{equation}}
\newcommand{\eeq}{\end{equation}}
\newcommand{\bea}{\begin{eqnarray}}
\newcommand{\eea}{\end{eqnarray}}

\newcommand{\GeV}{${\rm GeV}$}

\newcommand\TeV{\,\mbox{TeV}}

\graphicspath{/usr1/hboett/plots/qcd_pol/}
\graphicspath{./}

\begin{document}
\begin{titlepage}

\begin{flushleft}
DESY 10--184 %\hfill {\tt arXiv:10xx.yyyy [hep-ph]} 
\\
DO--TH 10/20\\
ZU-TH-17/10\\
SFB-CPP/10--118\\
November 2010 \\
%\today \\
\end{flushleft}

\vspace{1.5cm}
\noindent
\begin{center}
{\LARGE\bf \boldmath NNLO Benchmarks for Gauge and Higgs Boson}

\vspace{2mm}
{\LARGE\bf \boldmath Production at TeV Hadron Colliders}
\end{center}
\begin{center}

\vspace{2cm}

{\large S. Alekhin$^{a,b}$\footnote{Helmholtz Alliance Fellow}, J. Bl\"umlein$^a$, P. Jimenez-Delgado$^c$, 
S. Moch$^a$, and E. Reya$^d$}

\vspace{1.5cm}
{\it $^a$~Deutsches Elektronen--Synchrotron, DESY,\\
Platanenallee 6, D--15738 Zeuthen, Germany}\\

\vspace*{2mm}
{\it $^b$~Institute for High Energy Physics \\
    142281 Protvino, Moscow region, Russia}\\

\vspace*{2mm}
{\it $^c$~Institut f\"ur Theoretische Physik, Universit\"at Z\"urich, \\
Winterthurer Str. 190,
CH--8057 Z\"urich, Switzerland}\\

\vspace*{2mm}
{\it $^d$~Technische Universit\"at Dortmund, Institut f\"ur Physik, \\ D--44221 Dortmund, Germany}\\

%%\today
\vspace{1.5cm}
\end{center}

\begin{abstract}
\noindent
The inclusive production cross sections for $W^+, W^-$ and $Z^0$-bosons form important benchmarks
for the physics at hadron colliders. We perform a detailed comparison of the predictions for these 
standard candles based on recent next-to-next-to-leading order (NNLO) parton parameterizations and  new 
analyses including the combined HERA data, compare to all available experimental results, and discuss the 
predictions for present and upcoming RHIC, SPS, Tevatron and LHC energies. The rates for gauge boson 
production at the LHC can be rather confidently predicted with an accuracy of better than about 10\% at NNLO. We also 
present detailed NNLO predictions for the Higgs boson production cross 
sections for Tevatron and LHC energies (1.96, 7, 8, 14 TeV), and propose a possible method to monitor the 
gluon distribution experimentally in the kinematic region close to the mass range expected 
for the Higgs boson. The production cross sections of the Higgs boson at the LHC are presently predicted 
with an accuracy of about 10--17\%. The inclusion of the NNLO contributions is mandatory for achieving such 
accuracies since the total uncertainties are substantially larger at NLO. 
\end{abstract}
\end{titlepage}

\vfill
\newpage
\sloppy

%%%%%%%%%%%%%%%%%%%%%%%%%%%%%%%%%%%%%%%%%%%%%%%%%%%%%%%%%%%%%%%%%%%%%%%
%	Introduction
%%%%%%%%%%%%%%%%%%%%%%%%%%%%%%%%%%%%%%%%%%%%%%%%%%%%%%%%%%%%%%%%%%%%%%%
\noindent
{\large \sf {1~~Introduction.}}
%\label{sec:intro}
%
%\vspace{2mm}
%\noindent
~~The inclusive production cross sections of the weak gauge bosons $W^{\pm}, Z^0$ and of the 
Higgs particle $H^0$ form important reference points for the physics at hadron colliders.
Within QCD the corresponding cross sections have been calculated to  
NNLO \cite{WZNNLO,H0NNLO}.~\footnote{Electro--weak corrections were calculated in 
\cite{WZEW,H0EW,Anastasiou:2008tj}.}  This level of accuracy is necessary to control the renormalization 
and factorization scale uncertainties of the parton distribution functions (pdfs), which are still of significant
size at next-to-leading order (NLO). The rapidly growing luminosity at the Large Hadron Collider, LHC,
allows a precise measurement of these quantities \cite{CMS,ATLAS}, which reaches the accuracy of the
NNLO predictions, based on the world deep-inelastic, Drell--Yan and di-muon data, 
and part of the Tevatron data~\cite{ABM,ABKM,JR,HERAPDF,MSTW08,MSTW10}. These analyses also require a correct 
treatment of the heavy flavor contributions, cf.~\cite{HQ}, which yield sizeable effects.

In this note we provide NNLO predictions for the inclusive weak boson and Standard Model Higgs-boson 
production cross sections, and compare to the available experimental data 
\cite{UA1,UA2,PHENIX,CDF,CDF1,D0,CMS,ATLAS}. For present and upcoming experimental analyses we also provide 
detailed reference tables, and comment on the well-known NLO predictions, cf.~\cite{NLO}. These are, however, 
far less precise due to their inherent large renormalization and factorization scale uncertainties \cite{ALT,BRNV}.  

The effect of the new combined HERA data on the inclusive NNLO $W^\pm, Z^0$ and $H^0$ boson cross sections 
\cite{COMB} is shown by comparing the results for the recent ABM10 \cite{ABM} and ABKM09 
\cite{ABKM} distributions. It leads to 
significant
shifts of the total rate of about $1 \sigma$ in the pdf--error for a wide range of collider energies and 
processes.
We consider all collider energies having been probed so far, compare to  all measurements,
and give predictions for the high energy options at the LHC. In this way a wide kinematic region in Bjorken $x$ 
is probed for the corresponding parton luminosities which may help to delineate remaining  
differences in 
the current NNLO pdfs and to devise a way of potential further improvements.
%%%%%%%%%%%%%%%%%%%%%%%%%%%%%%%%%%%%%%%%%%%%%%%%%%%%%%%%%%%%%%%%%%%%%%%
%	W and Z
%%%%%%%%%%%%%%%%%%%%%%%%%%%%%%%%%%%%%%%%%%%%%%%%%%%%%%%%%%%%%%%%%%%%%%%

\vspace*{7mm}
\noindent
{\large \sf {2~~W$^\pm$ and Z$^0$ Boson Production.}}
%\label{sec:wz}
%
%\vspace{2mm}
%\noindent
~~The inclusive $W^\pm$ and $Z^0$ boson production cross sections in $p\overline{p}$ and $pp$ scattering are 
known 
to 2nd order (NNLO) in the strong coupling constant~\cite{WZNNLO}, supplemented by the 1st order (NLO)
electroweak corrections, cf.~\cite{WZEW}. In the following comparison we will concentrate on the QCD 
corrections only. The electroweak parameters are calculated choosing the scheme based on $(G_F, M_W, M_Z)$ 
\cite{PDG2010} with $G_F = 1.16637 \times 10^{-5}~\GeV^{-2}, M_W = 80.399  \pm 0.023~\GeV, 
M_Z = 91.1876  \pm 0.0021~\GeV,$ and the weak mixing angle as dependent quantity, with  
%-------------------------------------------------------------------------------------------------------------
\begin{equation}
\hat{s}_Z^2 =  1 - \frac{M_W^2}{\hat{\rho} M_Z^2}  = 0.2307 \pm 0.0005~,
\end{equation}
%-------------------------------------------------------------------------------------------------------------
and $\hat{\rho} = 1.01047 \pm 0.00015$. Furthermore, the width of the $W^\pm$ and $Z^0$ bosons are 
$\Gamma(W^\pm) = 2.085 \pm 0.042$ GeV, $\Gamma(Z^0) = 2.4952 \pm 0.0023$ GeV, and $\sin^2 \theta_c 
= 0.051$, with 
$\theta_c$ 
the Cabibbo angle. We compute the inclusive production cross sections at various collider energies
for the recent NNLO parton distributions, ABM10 
\cite{ABM}, ABKM09 \cite{ABKM}, JR \cite{JR}, MSTW08 \cite{MSTW08}, and HERAPDF \cite{HERAPDF}. In the latter 
case we refer to the fit where a value of $\alpha_s(M_Z^2) = 0.1145$ has been assumed at NNLO, which resulted 
in  the lowest 
value
for $\chi^2_{\rm min}$ among other choices.~\footnote{For the time being the NNLO HERAPDF parameterization yields only 
central values, unlike the NLO distributions, where also errors are provided, cf.~\cite{HERAPDF}.}  

In Table~\ref{tab:T1} we summarize the cross sections for the different $p\overline{p}$ collision energies at 
the SPS and the Tevatron at NNLO for the distributions \cite{ABM,ABKM,JR,MSTW08,HERAPDF} . For one of the 
parameterizations, JR, we compare also to the NLO QCD corrections. The relative differences between the NLO 
and NNLO corrections are of comparable size for all other pdf sets. We also list the corresponding NNLO values of 
$\alpha_s(M_Z^2)$ and their errors as determined, or partly being assumed, in the various analyses, which in 
most cases turn out to be similar. 
We note that the corresponding $\alpha_s$ values at NLO, despite differences in the central values, necessarily
all agree within the rather large theory errors due to factorization and renormalization scale 
uncertainties of at least $\pm 0.005$, cf. \cite{ALT,BRNV}. This also applies to other NLO analyses 
\cite{CTEQ,NNPDF}. 
%%%%%% Table 1 %%%%%%%%%%%%%%%%%%%%%%%
\restylefloat{table}
%\begin{landscape}
\begin{table}[H]  %%[h]
\small
\begin{center}
\renewcommand{\arraystretch}{1.5}
\begin{tabular}{|c|c|c|c|c|c|}
\hline
$\sqrt{s}$ (TeV) & & 0.546 & 0.630 & 1.8 & 1.96\\
\hline
ABM10 \cite{ABM}  & $W^\pm$ & 
   5.632 $\pm$ 0.092 & 
   7.045 $\pm$ 0.111 &   24.441 $\pm$ 0.235 & 
  26.740 $\pm$ 0.259 \\
$\alpha_s=0.1147 \pm 0.0012$
&  $Z^0$  & 
   1.761 $\pm$ 0.022 & 
   2.187 $\pm$ 0.028 &
   7.181 $\pm$ 0.068 & 
   7.846 $\pm$ 0.075 \\
\hline
ABKM09 \cite{ABKM} & $W^\pm$  &
5.804 $\pm$ 0.075   &
7.222 $\pm$ 0.091   &
23.88 $\pm$ 0.243   &
26.09 $\pm$ 0.265   \\
$\alpha_s=0.1135 \pm 0.0014$    &  $Z^0$  &
1.806 $\pm$ 0.020 &
2.234 $\pm$ 0.024 &
7.056 $\pm$ 0.068 & 
7.691 $\pm$ 0.075 \\
\hline
JR \cite{JR} & $W^{\pm}$  & 
   5.983 $\pm$ 0.148 
&  7.346 $\pm$ 0.159 
& 23.069 $\pm$ 0.238 
& 25.157 $\pm$ 0.251 \\
$\alpha_s= 0.1124\pm 0.0020$
& & (5.358 $\pm$ 0.152)
  & (6.637 $\pm$ 0.167)
  & (22.121 $\pm$ 0.274) 
  &  (24.181 $\pm$ 0.296)\\
%-----
& $Z^0$  & 1.837 $\pm$ 0.029 
         & 2.268 $\pm$ 0.034 
         & 6.975 $\pm$ 0.071 
         & 7.586 $\pm$ 0.076 \\
& &        (1.648 $\pm$ 0.028)  
&          (2.047 $\pm$ 0.033) 
&          (6.667 $\pm$ 0.080) 
&          (7.272 $\pm$ 0.087)\\
%--------------------------------------------------------------------
\hline
MSTW08 \cite{MSTW08}  & $W^\pm$ &
5.469 $\pm$ 0.151 & 
6.802 $\pm$ 0.176& 
23.14 $\pm$ 0.394&
25.35 $\pm$ 0.422\\
$\alpha_s= 0.1171 \pm 0.014$    &  $Z^0$  & 
1.654 $\pm$ 0.047&
2.056 $\pm$ 0.056&
6.773 $\pm$ 0.126&
7.406 $\pm$ 0.134
\\
%--------------------------------------------------------------------
\hline
HERAPDF \cite{HERAPDF} & $W^{\pm}$ & 
6.121 & 
7.519 & 
24.51 &
26.80 \\
$\alpha_s=0.1145 $& $Z^0$ & 
1.853 & 
2.296 & 
7.319 &
7.978 \\
\hline
\end{tabular}
%---------------------
\caption{\label{tab:T1}
\small NNLO predictions for the production cross sections $\sigma(p\overline{p} \to V+X)$ [$nb$], with 
$V=W^{\pm},Z^0$.  The abbreviation $W^\pm$ refers to the sum $W^+ +W^-$. Notice that for $p\overline{p}$
collisions the $W^+$ and $W^-$ cross sections are equal. The errors refer to the $\pm 1 
\sigma$ pdf uncertainties. The NNLO values of $\alpha_s$ refer to $\alpha_s=\alpha_s(M_Z^2)$. 
To allow for a comparison  with the corrections up to NLO  the corresponding cross sections  
for the JR distributions are also listed as an example in parentheses~\cite{GJR}.}
%---------------------
\end{center}
\end{table}
%\vfill
%\end{landscape}
%%%%%%%%%%%%%%%%%%%%%%%%%%%%%%%%
\noindent

The NNLO corrections furthermore {\it enhance} the cross sections at all center-of-mass 
(cms) energies both for $W^\pm$ and $Z^0$ boson production. 
The importance of full NNLO analyses becomes evident when comparing the NLO and NNLO predictions in 
Tables~\ref{tab:T1} 
and \ref{T2}.  At the Tevatron (1.96 TeV), for example, $\Delta\sigma^{W^{\pm}}\equiv \sigma_{\rm 
NNLO}^{W^{\pm}}-
\sigma_{\rm NLO}^{W^{\pm}}= + 0.930$, $+0.976$ and $+0.825$  $nb$ for ABM10, JR and MSTW08,
respectively. Similarly for LHC (7 TeV) one obtains $\Delta\sigma^{W^{\pm}} =+1.05$,  $+2.36$
and $+2.46$~$nb$ for ABM10, JR and MSTW08, respectively.  This corresponds at the Tevatron
to a difference of more than $3 \sigma$
w.r.t.\ the pdf--errors, and to more than $1\sigma$ at LHC.  Furthermore, at NNLO the parton distributions
are much more stable against renormalization and factorization scale uncertainties than at NLO :
the scale variations of the full NNLO predictions in Table~\ref{tab:T1} amount to less than 0.5 \% (i.e., are
about half as large as the stated  $1 \sigma_{\rm pdf}$ uncertainties) which is about {\it four times smaller} 
than at NLO \cite{GJR,JR}; at LHC energies the scale uncertainties of the NNLO predictions in Table~\ref{T2}
amount to less than 2 \% of the total predicted rates which is about {\it half} as large as the stated 
$1 \sigma_{\rm pdf}$ uncertainties and the scale uncertainties at NLO  \cite{GJR,JR}.

Thus far, the combined H1 and ZEUS data \cite{COMB} have been
taken into account in the ABM10 \cite{ABM} and HERAPDF \cite{HERAPDF} analyses only, and their relative effect
can be seen by comparing the numbers for the ABM10 and ABKM09 distributions at NNLO. While for the lower 
collider 
energies 2--3.5\% lower cross sections are obtained, about 2.5 \% larger cross sections result for the Tevatron 
energies of $1.8$ and $1.96$ TeV. With respect to the current pdf--errors this effect amounts to $2 \sigma$.
%%%%\clearpage
%%%%%%%Table 2 %%%%%%%%%%%%%%%%%%%%%%%
%\begin{landscape}
\begin{table}[H]
%\vspace*{-1.5cm}
\footnotesize
\begin{center}
\renewcommand{\arraystretch}{1.5}
\begin{tabular}{|c|c|c|c|c|c|}
\hline
$\sqrt{s}$ (TeV) & & 0.5 & 7 & 10 & 14\\
%-----------------------------------------------------------------
\hline
ABM10 \cite{ABM}  
%--
&  $W^+$  
&  1.236 $\pm$ 0.057
&  59.86 $\pm$ 0.838
&  85.58 $\pm$ 1.267
&  118.4 $\pm$ 1.891
\\
%--
& $W^{-}$ 
& 0.363 $\pm$ 0.092 
& 40.28 $\pm$ 0.535
& 60.28 $\pm$ 0.852
& 86.58 $\pm$ 1.331
\\
%--
& $W^\pm$  
& 1.600 $\pm$ 0.070
& 100.1 $\pm$ 1.315 
& 145.9 $\pm$ 2.065 
& 205.0 $\pm$ 3.186
%--
\\
& $Z^0$ 
& 0.305 $\pm$ 0.015 
& 29.01 $\pm$ 0.391
& 42.77 $\pm$ 0.633
& 60.69 $\pm$ 0.963
\\
%-----------------------------------------------------------------
\hline
ABKM09 \cite{ABKM}  
&  $W^+$  
&  1.160 $\pm$ 0.046
&  58.86 $\pm$ 0.903
&  85.14 $\pm$ 1.427
&  119.4 $\pm$ 2.072
\\
%--
& $W^-$  
& 0.348 $\pm$ 0.014
& 39.43 $\pm$ 0.614
& 59.56 $\pm$ 0.993
& 86.53 $\pm$ 1.525
\\
%--
& $W^{\pm}$ 
& 1.509 $\pm$ 0.058
& 98.27 $\pm$ 1.527
& 144.7 $\pm$ 2.436
& 205.9 $\pm$ 3.658
%--
\\
& $Z^0$ 
& 0.287 $\pm$ 0.012 
& 28.42 $\pm$ 0.457
& 42.28 $\pm$ 0.743 
& 60.70 $\pm$ 0.115
\\
%-----------------------------------------------------------------
\hline
JR \text{\cite{JR}} &
${W}^+$   & 1.138 $\pm$ 0.061 & 54.57 $\pm$ 1.10
&  78.43 $\pm$ 1.98 & 109.31
$\pm$ 3.13 \\
%--
& & (1.245 $\pm$ 0.065) & (52.96 $\pm$ 0.99) & (76.60 $\pm$ 1.74) & (107.58 $\pm$ 2.95) \\
%--
& ${W}^-$   & 0.387 $\pm$ 0.028 & 37.15 $\pm$ 0.79
&  55.54 $\pm$ 1.44  &  80.02
$\pm$ 2.31 \\
%--
&  & (0.427 $\pm$ 0.030) & 
(36.39 $\pm$ 0.72) & (54.67 $\pm$ 1.26) & (79.16 $\pm$ 2.12) \\
%--
& ${W}^\pm$ & 1.525 $\pm$ 0.052 & 91.72 $\pm$ 1.82
& 133.99 $\pm$ 3.35 & 189.29
$\pm$ 5.41 \\
%--
& & (1.672 $\pm$ 0.053) & (89.36 $\pm$ 1.57) & (131.23 $\pm$ 2.87) & (186.74 $\pm$ 4.95) \\
%--
& ${Z}^0$   & 0.300 $\pm$ 0.011 & 27.24 $\pm$ 0.50 &  40.39 $\pm$ 0.95 &  57.85
$\pm$ 1.56 \\
%--
& & (0.336 $\pm$ 0.012) & (26.57 $\pm$ 0.43) &  (39.57 $\pm$ 0.81) &  (57.00 $\pm$ 1.42) \\
%-----------------------------------------------------------------
\hline
MSTW08 \cite{MSTW08,MSTW10} 
& $W^+$ 
& 1.221 $\pm$ 0.0421
& 56.80 $\pm$ 0.971
& 81.83 $\pm$ 1.405 
& 114.0 $\pm$ 1.945
\\
%---
&  $W^-$  
&0.416 $\pm$ 0.017
&39.63 $\pm$ 0.678
&59.45 $\pm$ 1.008
&85.63 $\pm$ 1.484
\\
%---
& $W^{\pm}$ 
& 1.637 $\pm$ 0.052
& 96.41 $\pm$ 1.607
& 141.3 $\pm$ 2.372 
& 199.6 $\pm$ 3.379
\\
%---
& $Z^0$ 
& 0.319 $\pm$ 0.011
& 27.89 $\pm$ 0.481
& 41.34 $\pm$ 0.705
& 58.99 $\pm$ 1.012
\\
%-----------------------------------------------------------------
\hline
HERAPDF \cite{HERAPDF} 
& $W^+$ 
& 1.219
& 59.37
& 85.37
& 119.0
\\
%--
&  $W^-$  
& 0.414
& 40.82  
& 61.06
& 87.94
\\
%--
& $W^{\pm}$ 
& 1.633
& 100.2
& 146.4
& 206.9 
\\
%--
& $Z^0$ 
& 0.322
& 29.08 
& 42.95
& 61.22
\\
%-----------------------------------------------------------------
\hline
\end{tabular}
%---------------------
\caption{\label{T2}
\small NNLO predictions for the production cross sections $\sigma(pp \to V+X)$ [$nb$], with 
$V=W^{\pm},Z^0$. The abbreviation $W^\pm$ denotes the sum $W^+ + W^-$.  
The errors refer to the $\pm 1 \sigma$ pdf uncertainties. 
To allow for a comparison  with the corrections up to NLO  we also listed the corresponding 
cross sections for the JR distributions as an example in parentheses \cite{GJR}.}
%---------------------
\end{center}
\end{table}
%\vfill
%\end{landscape}
%%%%%%%%%%%%%%%%%%%%%%%%%%%%%%%%

For illustration we compare
in Figure~1(a) the different predictions \cite{ABM,JR,MSTW08,HERAPDF} with each other 
and with the measured cross sections \cite{UA1,UA2,PHENIX,CDF,CDF1,D0} at fixed energies in the range 
$\sqrt{S} = 
0.5 - 1.96$ TeV, covering also the $pp$ cross sections at RHIC, cf. Table~\ref{T2}.  A detailed summary 
of the various measurements is given in the Appendix.
As can also be seen in Table~\ref{tab:T1} the 
results for $Z^0$ production predicted by the different parameterizations
agree somewhat
better than in the case of $W^\pm$ production. The experimental errors
at lower energies $\sqrt{S} < 0.63$ TeV are rather large and the current NNLO predictions agree with 
experiment. For $\sqrt{S} = 0.63$ TeV all NNLO analyses predict cross sections which are at the upper end 
of the measurements. At Tevatron energies the ABM10 and HERAPDF distributions yield about 5--7\% larger 
$W^\pm$ cross sections than those of JR and MSTW08. 
Moving to higher energies the JR distributions result in lower cross sections as compared 
to the ABM10 and HERAPDF distributions, which yield larger values at lower energies as well. Yet all 
predictions are in agreement with the current experimental results at the level of $1.5 \sigma$.

\restylefloat{figure}
\begin{center}
\begin{figure}[H] %%[b]
\epsfig{figure=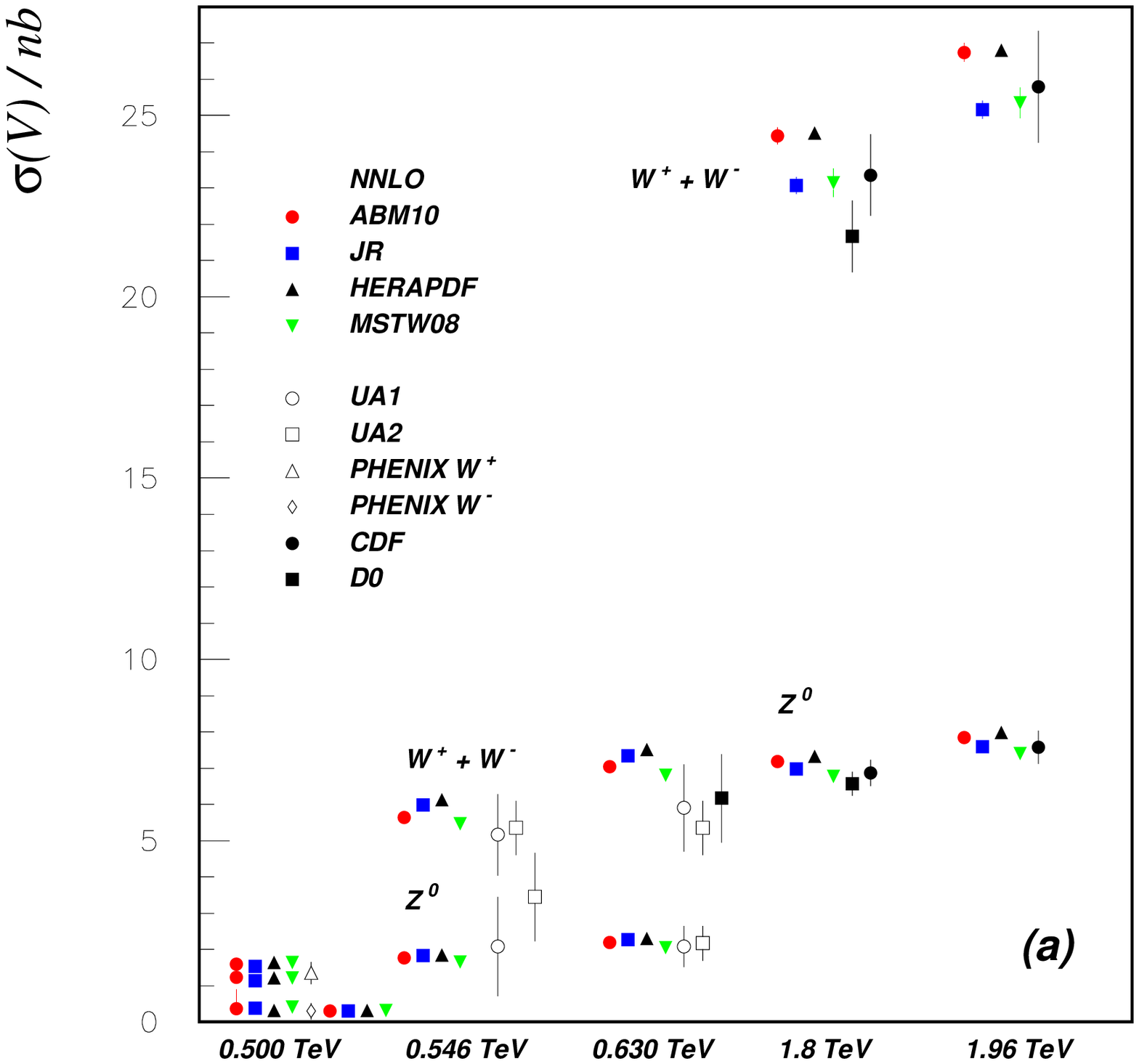,height=10cm,width=0.48\linewidth} \hspace*{2mm}
\epsfig{figure=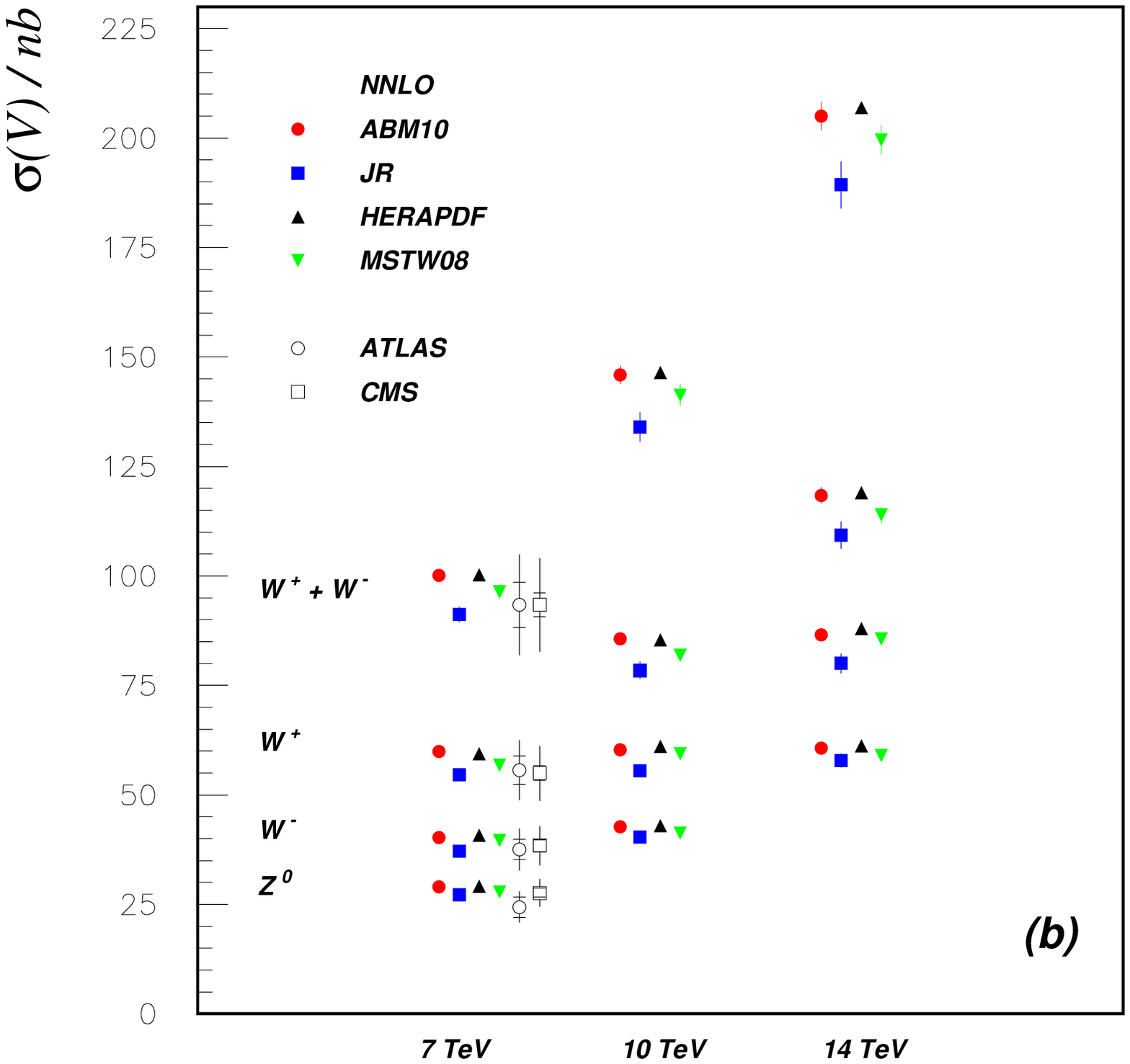,height=10cm,width=0.48\linewidth} 
\caption[]{
\label{FIG:WZ}
\small Comparison of different NNLO predictions for the inclusive $W^+$, $W^-$, $W^\pm$, and $Z^0$ 
boson production cross sections in $p\overline{p}$ annihilation and $pp$ scattering ($\sqrt{S} = 0.5$ TeV) 
based on the pdfs of recent NNLO analyses \cite{ABM,ABKM,JR,HERAPDF,MSTW08,MSTW10} and the corresponding
experimental data \cite{UA1,UA2,PHENIX,CDF,CDF1,D0,ATLAS,CMS}. Left panel (a): the lower energy region 
corresponds to $p\overline{p}$ collisions, except at 0.5 TeV, which refers to $pp$ scattering.
For the latter case the predictions refer to (from above) $W^+ + W^-, W^+, W^-$ and the ones for $Z^0$ are
given to the right of the ones for $W^-$.
Right 
panel (b): LHC energies ($pp$ collisions); the inner error bars refer to $(\sigma_{\rm stat}^2+\sigma_{\rm 
syst}^2)^{1/2}$ and the total error is obtained by adding the luminosity error in quadrature.}
\end{figure}
\end{center}

\vspace*{-1.3cm}
In Table~\ref{T2} we list the NNLO predictions for the $pp$ production cross sections for $W^+$, $W^-$ bosons, 
their sum, as well as those 
for $Z^0$ bosons at RHIC and for different present or planned collider energies at  the LHC.
As before we supplement the NLO prediction for one set of distributions (JR) 
to allow for comparisons as an example. For the LHC energies the NNLO corrections lead to an enhancement
of the cross sections w.r.t. NLO, while at RHIC energies the NNLO cross sections become smaller. At LHC 
energies the pattern of relative differences of the various predictions 
\cite{ABM,ABKM,JR,MSTW08,HERAPDF,MSTW10} 
both for $W^\pm$ and $Z^0$ boson 
production pertains  over the whole energy range, see also Figure~1(b). The cross sections grow 
empirically nearly linearly with $\sqrt{S}$. The impact of the combined HERA data is less than at lower
collider energies and leads in most cases only to a very slight enhancement of the cross sections, as a comparison of the
values in Table~\ref{T2} resulting from the ABM10 and ABKM09 distributions shows. The largest scattering cross 
sections are 
obtained for the ABM10 and HERAPDF distributions, followed by slightly lower values for MSTW08, and even
somewhat lower cross sections are obtained for the JR distributions, with differences of up to 9\%.
Comparing to the first experimental measurements by CMS \cite{CMS} and ATLAS \cite{ATLAS}, all current 
predictions
are compatible within the present experimental errors for all channels.

Due to the reduced scale uncertainty at NNLO and the slightly different NNLO estimates of the various groups, cf. 
Figures~1(a) and (b), and Tables \ref{tab:T1} and \ref{T2}, we conclude that the rates for gauge boson 
production at LHC energies can be rather confidently predicted with an accuracy better than about 9\%. These 
differences are due to the different light sea distributions  ($\bar{u}, \bar{d}, \bar{s}$) obtained in the various analyses and 
require  more 
detailed
theoretical and experimental investigations in the future.  
%%%%%%%%%%%%%%%%%%%%%%%%%%%%%%%%%%%%%%%%%%%%%%%%%%%%%%%%%%%%%%%%%%%%%%%
%	Higgs
%%%%%%%%%%%%%%%%%%%%%%%%%%%%%%%%%%%%%%%%%%%%%%%%%%%%%%%%%%%%%%%%%%%%%%%

\vspace*{5mm}
\noindent
{\large \sf {3~~Higgs Boson Production.}}
%\section{\boldmath Higgs Boson Production}
\label{sec:hi}
%
%\vspace{2mm}
%\noindent
~~The inclusive partonic production cross section~\footnote{The vector boson fusion channel adds a 
correction of about $10\%$, with an $O(2 \%)$ pdf uncertainty, cf.~\cite{MB}.} 
of the Higgs boson of the Standard Model $\sigma(pp \rightarrow H^0+X)$ \cite{H0NNLO}  
depends on the strong 
coupling constant $\propto \alpha_s^2$ and a large part of the cross section exhibits a quadratic dependence of 
the gluon 
density. As has been shown in Table~\ref{tab:T1}, the $\alpha_s(M_Z^2)$ values determined, resp. assumed 
\cite{HERAPDF}, are 
still different. 
Despite having reached an impressive accuracy of $O(1\%)$ in individual analyses of the world 
deep-inelastic data and related collider data, \cite{ABM,ABKM,JR,Martin:2009bu,BBG}~\footnote{For a 
compilation see 
Ref.~\cite{Blumlein:2010rn}.},  and for a large number of other high energy processes and decay widths, 
cf.~\cite{Bethke:2009jm}, an overall agreement has not yet been reached.  As well-known, the gluon density 
and 
the 
value of $\alpha_s(M_Z^2)$ are anticorrelated, see e.g. the tables of 
%%%%%% Table HIGGS_1 %%%%%%%%%%%%%%%%%%%%%%%
%\begin{landscape}
\begin{table}[H]
\footnotesize
\begin{center}
\renewcommand{\arraystretch}{1.5}
\begin{tabular}{|c|c|c|c|c|c|}
\hline
$M_H$ (GeV) 
& ABM10   \text{\cite{ABM}}
& ABKM09  \text{\cite{ABKM}}
& JR    \text{\cite{JR}} 
& MSTW08  \text{\cite{MSTW08}}
& HERAPDF \text{\cite{HERAPDF}} \\
\hline
 100  & 1.438   $\pm$   0.066  & 1.380   $\pm$   0.076  & 1.593   $\pm$   0.091  & 1.682   $\pm$   0.046  & 1.417 \\
 110  & 1.051   $\pm$   0.052  & 1.022   $\pm$   0.061  & 1.209   $\pm$   0.078  & 1.265   $\pm$   0.038  & 1.055 \\
 115  & 0.904   $\pm$   0.047  & 0.885   $\pm$   0.055  & 1.060   $\pm$   0.072  & 1.104   $\pm$   0.034  & 0.917 \\
 120  & 0.781   $\pm$   0.042  & 0.770   $\pm$   0.050  & 0.933   $\pm$   0.067  & 0.968   $\pm$   0.031  & 0.800 \\
 125  & 0.677   $\pm$   0.038  & 0.672   $\pm$   0.045  & 0.823   $\pm$   0.062  & 0.851   $\pm$   0.029  & 0.700 \\
 130  & 0.588   $\pm$   0.034  & 0.589   $\pm$   0.041  & 0.729   $\pm$   0.058  & 0.752   $\pm$   0.026  & 0.615 \\
 135  & 0.513   $\pm$   0.031  & 0.518   $\pm$   0.037  & 0.647   $\pm$   0.054  & 0.666   $\pm$   0.024  & 0.541 \\
 140  & 0.449   $\pm$   0.028  & 0.456   $\pm$   0.034  & 0.576   $\pm$   0.050  & 0.591   $\pm$   0.022  & 0.479 \\
 145  & 0.394   $\pm$   0.025  & 0.403   $\pm$   0.031  & 0.514   $\pm$   0.047  & 0.527   $\pm$   0.020  & 0.424 \\
 150  & 0.347   $\pm$   0.023  & 0.358   $\pm$   0.028  & 0.461   $\pm$   0.044  & 0.471   $\pm$   0.018  & 0.377 \\
 155  & 0.306   $\pm$   0.020  & 0.318   $\pm$   0.026  & 0.413   $\pm$   0.041  & 0.421   $\pm$   0.017  & 0.336 \\
 160  & 0.271   $\pm$   0.019  & 0.283   $\pm$   0.024  & 0.371   $\pm$   0.039  & 0.378   $\pm$   0.016  & 0.300 \\
 165  & 0.240   $\pm$   0.017  & 0.253   $\pm$   0.022  & 0.335   $\pm$   0.036  & 0.341   $\pm$   0.014  & 0.269 \\
 170  & 0.213   $\pm$   0.015  & 0.226   $\pm$   0.020  & 0.302   $\pm$   0.034  & 0.307   $\pm$   0.013  & 0.241 \\
 175  & 0.190   $\pm$   0.014  & 0.203   $\pm$   0.019  & 0.274   $\pm$   0.032  & 0.278   $\pm$   0.012  & 0.217 \\
 180  & 0.169   $\pm$   0.013  & 0.182   $\pm$   0.017  & 0.248   $\pm$   0.030  & 0.251   $\pm$   0.012  & 0.195 \\
 185  & 0.151   $\pm$   0.012  & 0.164   $\pm$   0.016  & 0.225   $\pm$   0.028  & 0.228   $\pm$   0.011  & 0.176 \\
 190  & 0.136   $\pm$   0.011  & 0.148   $\pm$   0.015  & 0.205   $\pm$   0.027  & 0.207   $\pm$   0.010  & 0.159 \\
 200  & 0.109   $\pm$   0.009  & 0.121   $\pm$   0.013  & 0.170   $\pm$   0.024  & 0.172   $\pm$   0.009  & 0.131 \\
\hline
\end{tabular}
\caption{\label{T3}
\small 
NNLO predictions for the production cross sections $\sigma(p\overline{p} \to H^0+X)$ [$pb$] at $\sqrt{S} = 
1.96$
TeV. The errors refer to the $\pm 1 \sigma$ pdf uncertainties.}  
\end{center}
\end{table}
%\end{landscape}
%%%%%%%%%%%%%%%%%%%%%%%%%%%%%%%%

\noindent
the correlation coefficients given in 
\cite{ABKM}. 
The values of $\alpha_s(M_Z^2)$ determined in global analyses to derive the parton distribution 
functions 
have all lower 
central values than the world average in \cite{Bethke:2009jm}\footnote{For other high energy measurements 
of $\alpha_s(M_Z^2)$ yielding lower values see e.g. \cite{LOWA}.}, 
with a spread of $\Delta 
\alpha_s(M_Z^2) = 0.0047$. 
If one extended the uncertainty referring to the average $\alpha_s(M_Z^2)$-value for the $\tau$-decay 
\cite{Bethke:2009jm} 
one would obtain $\Delta \alpha_s(M_Z^2) = 0.0073$. Both uncertainties imply
%--------------------------------------------------------------------------------------------------------------------
\begin{eqnarray}
\label{EQ:ALS}
\Delta \sigma(p\overline{p}(p) \rightarrow H^0+X) \sim \left(\frac{\Delta \alpha_s}{\alpha_s}\right)^2 = 
8.4~\text{resp.}~13~\%~.
\end{eqnarray}
%--------------------------------------------------------------------------------------------------------------------

\begin{center}
\begin{figure}[t]
\epsfig{figure=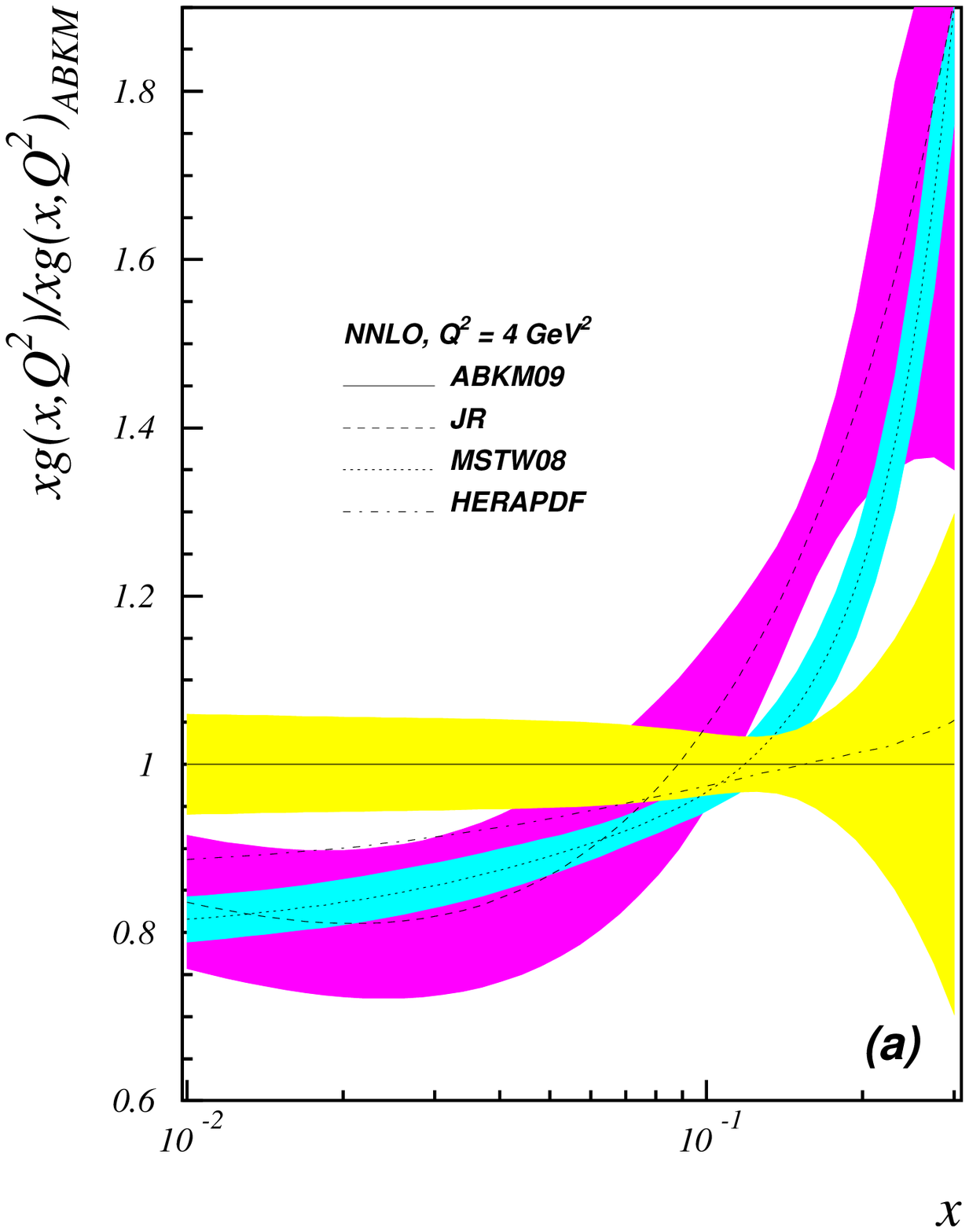,width=0.48\linewidth} \hspace*{3mm}
\epsfig{figure=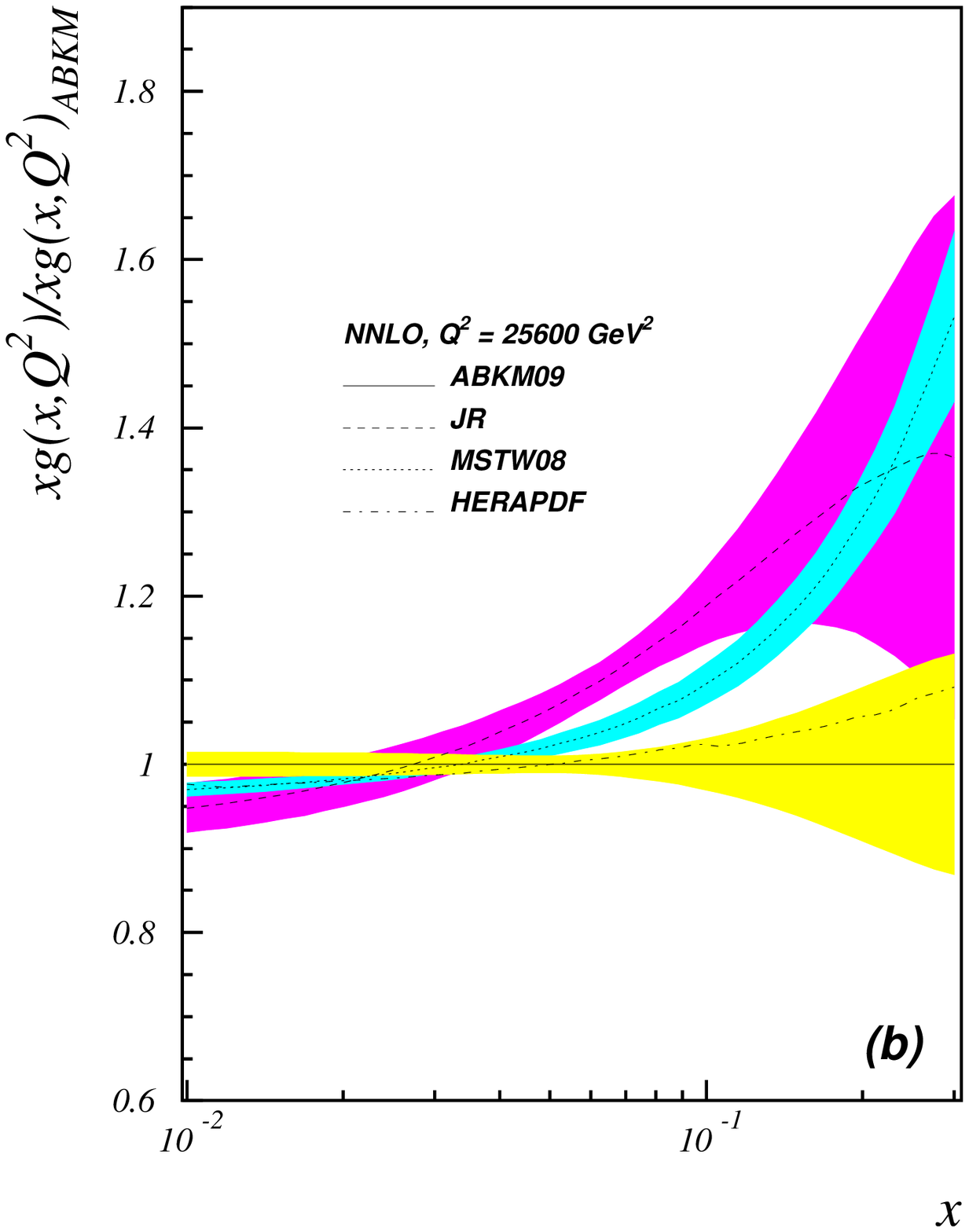,width=0.48\linewidth} 
\caption[]{\label{FIG:gluon}
\small Comparison of the NNLO gluon distributions at $Q^2 = 4~\GeV^2$ and $Q^2 = (160~\GeV)^2$
for the ratios $xg(x,Q^2)/xg(x,Q^2)_{\rm ABKM}$ 
for ABKM09 \cite{ABKM} (full line), JR \cite{JR} (dashed line),  
MSTW08 \cite{MSTW08} (dotted line), and HERAPDF \cite{HERAPDF} (dash-dotted line, without error band).} 
\end{figure}
\end{center}
\vspace{5mm}
\noindent

\vspace*{-1.2cm}
Furthermore, there are also still significant differences in the gluon distributions of different analyses.
The non-perturbative parton distributions at the initial scale $Q_0^2$ of the evolution 
are orthogonal in parameter space to $\Lambda_{\rm QCD}$, resp. $\alpha_s(M_Z^2)$, despite correlations.
The evolution is therefore a direct mapping of the initial conditions and its  strength is related
to the value of $\alpha_s(M_Z^2)$. The larger this parameter is, the faster the evolution. 
In Figure~2 we compare the NNLO gluon distributions of ABMKM09, JR, HERAPDF, and MSTW08 at
a scale $Q^2 = 4~\GeV^2$ and  $Q^2 = (160~\GeV)^2$, a typical mass scale for current Higgs 
boson 
searches, in the relevant $x$--range. Average values of Bjorken $\langle x \rangle \sim 10^{-1}$,
$\langle x \rangle = \sqrt{x_1 x_2} = M_H/\sqrt{S}$, in Figure~2(b) correspond to the production region at 
Tevatron, 
while those at $\langle x \rangle \sim 10^{-2}~(3 \times 10^{-2})$ are characteristic for the LHC at $\sqrt{S} 
= 14~(7) \TeV$. At $Q^2 = 4~\GeV^2$ the gluon distributions of JR and MSTW08 agree rather well in the region
$x \in [10^{-2}, 10^{-1}]$, while ABM10 and HERAPDF yield 20\% and 10\% larger values at $x = 
10^{-2}$.
All distributions get close for $x \sim 10^{-1}$ and for higher values the JR distributions becomes largest,
followed by MSTW08 and HERAPDF. The evolution to $\mu^2 = (160~\GeV)^2$ diminishes the differences at $x 
\sim 10^{-2}$ overall to about 5 \%.  All distributions cross around $x \sim 3 \times 10^{-2}$, and at $x \sim  
10^{-1}$ the gluon distributions by JR (MSTW08) take $\sim 20 \%~(10 \%)$ larger values than ABKM09, while the
HERAPDF values are close to the latter. Note that JR obtains a 4 \% smaller value of $\alpha_s(M_Z^2)$ than 
MSTW08, and  the value of $\alpha_s(M_Z^2)$ determined by ABKM is 3 \% smaller than that of MSTW08.  
%%%%%% Table HIGGS_2 %%%%%%%%%%%%%%%%%%%%%%%
%\begin{landscape}
\begin{table}[H]
\footnotesize
\begin{center}
\renewcommand{\arraystretch}{1.5}
\begin{tabular}{|c|c|c|c|c|c|}
\hline
$M_H$ (GeV) 
& ABM10   \text{\cite{ABM}}
& ABKM09  \text{\cite{ABKM}}
& JR    \text{\cite{JR}} 
& MSTW08  \text{\cite{MSTW08}}
& HERAPDF \text{\cite{HERAPDF}} \\
\hline
 100  & 22.82   $\pm$    0.53  & 21.18   $\pm$    0.60  & 20.48   $\pm$    0.70  & 22.95   $\pm$    0.31  & 20.90 \\
 110  & 18.65   $\pm$    0.44  & 17.30   $\pm$    0.49  & 16.92   $\pm$    0.56  & 18.84   $\pm$    0.26  & 17.12 \\
 115  & 16.95   $\pm$    0.40  & 15.72   $\pm$    0.45  & 15.46   $\pm$    0.50  & 17.16   $\pm$    0.23  & 15.58 \\
 120  & 15.45   $\pm$    0.37  & 14.34   $\pm$    0.41  & 14.17   $\pm$    0.45  & 15.69   $\pm$    0.22  & 14.22 \\
 125  & 14.14   $\pm$    0.35  & 13.12   $\pm$    0.38  & 13.03   $\pm$    0.41  & 14.39   $\pm$    0.20  & 13.03 \\
 130  & 12.96   $\pm$    0.32  & 12.03   $\pm$    0.35  & 12.01   $\pm$    0.37  & 13.23   $\pm$    0.19  & 11.97 \\
 135  & 11.92   $\pm$    0.29  & 11.07   $\pm$    0.33  & 11.10   $\pm$    0.34  & 12.20   $\pm$    0.17  & 11.02 \\
 140  & 10.99   $\pm$    0.27  & 10.21   $\pm$    0.31  & 10.29   $\pm$    0.32  & 11.28   $\pm$    0.16  & 10.18 \\
 145  & 10.15   $\pm$    0.26  &  9.44   $\pm$    0.29  &  9.55   $\pm$    0.29  & 10.45   $\pm$    0.15  &  9.42 \\
 150  &  9.40   $\pm$    0.24  &  8.75   $\pm$    0.27  &  8.89   $\pm$    0.27  &  9.71   $\pm$    0.14  &  8.74 \\
 155  &  8.73   $\pm$    0.23  &  8.13   $\pm$    0.25  &  8.30   $\pm$    0.25  &  9.04   $\pm$    0.14  &  8.13 \\
 160  &  8.12   $\pm$    0.21  &  7.56   $\pm$    0.24  &  7.75   $\pm$    0.24  &  8.43   $\pm$    0.13  &  7.57 \\
 165  &  7.56   $\pm$    0.20  &  7.05   $\pm$    0.23  &  7.26   $\pm$    0.23  &  7.88   $\pm$    0.12  &  7.07 \\
 170  &  7.06   $\pm$    0.19  &  6.59   $\pm$    0.21  &  6.82   $\pm$    0.21  &  7.38   $\pm$    0.12  &  6.62 \\
 175  &  6.60   $\pm$    0.18  &  6.17   $\pm$    0.20  &  6.41   $\pm$    0.20  &  6.92   $\pm$    0.11  &  6.20 \\
 180  &  6.19   $\pm$    0.17  &  5.79   $\pm$    0.19  &  6.04   $\pm$    0.19  &  6.51   $\pm$    0.11  &  5.83 \\
 185  &  5.80   $\pm$    0.16  &  5.43   $\pm$    0.18  &  5.70   $\pm$    0.18  &  6.13   $\pm$    0.10  &  5.48 \\
 190  &  5.46   $\pm$    0.15  &  5.11   $\pm$    0.17  &  5.39   $\pm$    0.18  &  5.78   $\pm$    0.10  &  5.16 \\
 200  &  4.84   $\pm$    0.14  &  4.55   $\pm$    0.16  &  4.83   $\pm$    0.16  &  5.16   $\pm$    0.09  &  
4.60 \\
 220  &  3.88   $\pm$    0.12  &  3.67   $\pm$    0.14  &  3.96   $\pm$    0.14  &  4.20   $\pm$    0.08  &  3.73 \\
 240  &  3.18   $\pm$    0.10  &  3.02   $\pm$    0.12  &  3.32   $\pm$    0.13  &  3.49   $\pm$    0.07  &  3.09 \\
 260  &  2.66   $\pm$    0.09  &  2.55   $\pm$    0.10  &  2.84   $\pm$    0.12  &  2.96   $\pm$    0.06  &  2.61 \\
 280  &  2.28   $\pm$    0.08  &  2.19   $\pm$    0.09  &  2.48   $\pm$    0.11  &  2.58   $\pm$    0.06  &  2.26 \\
 300  &  2.00   $\pm$    0.08  &  1.94   $\pm$    0.09  &  2.23   $\pm$    0.11  &  2.29   $\pm$    0.06  &  2.00 \\
\hline
\end{tabular}
\normalsize
\caption{\label{T4}
\small
NNLO predictions for the production cross sections $\sigma(pp \to H^0+X)$ [$pb$] at LHC for $\sqrt{S} = 
7$~TeV.  The errors refer to the $\pm 1 \sigma$ pdf uncertainties.}  
\end{center}
\end{table}
%\vfill
%\end{landscape}
%%%%%%%%%%%%%%%%%%%%%%%%%%%%%%%%

In Table~\ref{T3} the predictions for the NNLO Higgs boson production cross section for $p\overline{p}$ 
annihilation at $\sqrt{S} = 1.96$ TeV are compared for the distributions 
\cite{ABM,ABKM,JR,HERAPDF,MSTW08}\footnote{See also \cite{BD}.} for the mass range $100 \leq M_H \leq 
200$ GeV.  Here the largest cross 
section differences are those between MSTW08 and ABKM09 which amount to +22\% at $M_H = 100$ GeV and 
the MSTW08 prediction is +39\% 
higher than the one of ABM10 in the present exclusion region around $M_H = 160$ GeV. This 
difference corresponds 
to $5.6 \sigma$ in the pdf--error and is due to both the different gluon densities and $\alpha_s$ values 
having been obtained 
in both analyses, as discussed above. The impact of the combined H1 and ZEUS data is seen by comparing
the cross  sections for ABM10 and ABKM09~:
Larger cross sections are obtained at low masses $M_H \leq 140$ GeV, 
while for higher masses the cross sections become smaller than the ones by ABKM09. The cross sections predicted by JR and MSTW08 
are 
compatible within $1 \sigma$ and those of HERAPDF are about $1.5 \sigma$ larger than those of ABM10. From the 
above discussion of the gluon distribution and the different values of $\alpha_s(M_Z^2)$ involved, these
quantitative relations can be understood to a large extent. To establish firm exclusion bounds, 
cf.~\cite{HWGR:2010ar}, 
the observed variation in the predicted cross sections has to be taken into account as an uncertainty.

In Tables~\ref{T4} and \ref{T5} we compare the different NNLO predictions for $pp$--scattering at the 
current LHC energy of 7~TeV and the anticipated one of 8~TeV for the next running period \cite{RH} 
for the mass range of $100 \leq M_H \leq 300$ GeV. A quite similar pattern is obtained for both cms energies.
%%%%%% Table HIGGS_2 %%%%%%%%%%%%%%%%%%%%%%%
%\begin{landscape}
\begin{table}[H]
\footnotesize
\begin{center}
\renewcommand{\arraystretch}{1.5}
\begin{tabular}{|c|c|c|c|c|c|}
\hline
$M_H$ (GeV) 
& ABM10   \text{\cite{ABM}}
& ABKM09  \text{\cite{ABKM}}
& JR    \text{\cite{JR}} 
& MSTW08  \text{\cite{MSTW08}}
& HERAPDF \text{\cite{HERAPDF}} \\
\hline
 100  & 28.81   $\pm$    0.65  & 26.81   $\pm$    0.74  & 25.66   $\pm$    0.91  & 28.85   $\pm$    0.38  & 26.38 \\
 110  & 23.71   $\pm$    0.54  & 22.04   $\pm$    0.61  & 21.31   $\pm$    0.72  & 23.83   $\pm$    0.32  & 21.74 \\
 115  & 21.62   $\pm$    0.49  & 20.09   $\pm$    0.56  & 19.53   $\pm$    0.65  & 21.77   $\pm$    0.29  & 19.85 \\
 120  & 19.78   $\pm$    0.46  & 18.38   $\pm$    0.51  & 17.95   $\pm$    0.59  & 19.96   $\pm$    0.27  & 18.18 \\
 125  & 18.15   $\pm$    0.42  & 16.86   $\pm$    0.48  & 16.55   $\pm$    0.53  & 18.35   $\pm$    0.25  & 16.70 \\
 130  & 16.70   $\pm$    0.39  & 15.52   $\pm$    0.44  & 15.29   $\pm$    0.49  & 16.93   $\pm$    0.23  & 15.39 \\
 135  & 15.41   $\pm$    0.36  & 14.32   $\pm$    0.40  & 14.17   $\pm$    0.44  & 15.65   $\pm$    0.21  & 14.21 \\
 140  & 14.25   $\pm$    0.34  & 13.24   $\pm$    0.38  & 13.16   $\pm$    0.41  & 14.51   $\pm$    0.20  & 13.16 \\
 145  & 13.21   $\pm$    0.32  & 12.28   $\pm$    0.36  & 12.26   $\pm$    0.37  & 13.48   $\pm$    0.19  & 12.22 \\
 150  & 12.27   $\pm$    0.30  & 11.41   $\pm$    0.33  & 11.44   $\pm$    0.35  & 12.55   $\pm$    0.18  & 11.37 \\
 155  & 11.42   $\pm$    0.28  & 10.63   $\pm$    0.31  & 10.69   $\pm$    0.32  & 11.71   $\pm$    0.17  & 10.60 \\
 160  & 10.66   $\pm$    0.26  &  9.92   $\pm$    0.29  & 10.02   $\pm$    0.30  & 10.96   $\pm$    0.16  &  9.90 \\
 165  &  9.96   $\pm$    0.25  &  9.28   $\pm$    0.27  &  9.41   $\pm$    0.28  & 10.27   $\pm$    0.15  &  9.27 \\
 170  &  9.33   $\pm$    0.23  &  8.69   $\pm$    0.26  &  8.85   $\pm$    0.27  &  9.64   $\pm$    0.14  &  8.69 \\
 175  &  8.75   $\pm$    0.22  &  8.15   $\pm$    0.25  &  8.34   $\pm$    0.25  &  9.06   $\pm$    0.14  &  8.17 \\
 180  &  8.22   $\pm$    0.21  &  7.67   $\pm$    0.24  &  7.88   $\pm$    0.24  &  8.54   $\pm$    0.13  &  7.69 \\
 185  &  7.73   $\pm$    0.20  &  7.22   $\pm$    0.23  &  7.45   $\pm$    0.23  &  8.06   $\pm$    0.12  &  7.25 \\
 190  &  7.29   $\pm$    0.19  &  6.81   $\pm$    0.21  &  7.06   $\pm$    0.22  &  7.62   $\pm$    0.12  &  6.85 \\
 200  &  6.51   $\pm$    0.18  &  6.09   $\pm$    0.20  &  6.36   $\pm$    0.20  &  6.84   $\pm$    0.11  &  6.14 \\
 220  &  5.28   $\pm$    0.15  &  4.96   $\pm$    0.17  &  5.26   $\pm$    0.17  &  5.61   $\pm$    0.10  &  5.02 \\
 240  &  4.37   $\pm$    0.13  &  4.13   $\pm$    0.15  &  4.44   $\pm$    0.15  &  4.70   $\pm$    0.09  &  4.19 \\
 260  &  3.70   $\pm$    0.12  &  3.51   $\pm$    0.13  &  3.83   $\pm$    0.14  &  4.03   $\pm$    0.08  &  3.58 \\
 280  &  3.20   $\pm$    0.10  &  3.05   $\pm$    0.12  &  3.38   $\pm$    0.14  &  3.53   $\pm$    0.07  &  3.13 \\
 300  &  2.83   $\pm$    0.10  &  2.72   $\pm$    0.11  &  3.05   $\pm$    0.13  &  3.17   $\pm$    0.07  &  2.79 \\
\hline
\end{tabular}
\normalsize
\caption{\label{T5}
\small
NNLO predictions for the production cross sections $\sigma(pp \to H^0+X)$ [$pb$] at LHC for $\sqrt{S} = 8$~TeV.  
The errors refer to the $\pm 1 \sigma$ pdf uncertainties.}  
\end{center}
\end{table}
%\vfill
%\end{landscape}
%%%%%%%%%%%%%%%%%%%%%%%%%%%%%%%%
\noindent
At $\sqrt{S} = 7$ TeV the predictions of HERAPDF are lower than those of ABM10, while JR yields also smaller 
values 
for masses $M_H < 200$ GeV and larger ones for higher Higgs masses. MSTW08 predicts higher cross sections than 
ABM10, with a tendency of  a growing difference towards high masses. Overall the predictions are at variance
of up to $3 \sigma$ of the pdf--errors, which corresponds to maximal deviations of 11--14~\%. In the 
relevant region the gluon densities agree better than 5~\%, but there is still also the uncertainty 
in $\alpha_s(M_Z^2)$, see Eq.~(\ref{EQ:ALS}). At $\sqrt{S} = 8$ TeV the differences amount to less than 
$3.5 \sigma$ of the pdf--error, which corresponds to deviations of 11--16~\%.
In Table~\ref{T6} we compare the NNLO predictions for $\sqrt{S} = 14$ TeV for  Higgs masses between $100$ and 
$300$ GeV. They differ between 10--14\% and agree better for larger masses. These differences form theory
errors, which have to be accounted for within feasibility studies, cf. \cite{ATLAS:HIG}, and in searching for
Higgs boson production at the LHC.
%%%%%% Table HIGGS_3 %%%%%%%%%%%%%%%%%%%%%%%
%\begin{landscape}
\begin{table}[H]
\footnotesize
\begin{center}
\renewcommand{\arraystretch}{1.5}
\begin{tabular}{|c|c|c|c|c|c|}
\hline
$M_H$ (GeV) 
& ABM10   \text{\cite{ABM}}
& ABKM09  \text{\cite{ABKM}}
& JR   \text{\cite{JR}} 
& MSTW08  \text{\cite{MSTW08}}
& HERAPDF \text{\cite{HERAPDF}} \\
\hline
 100  & 71.16   $\pm$    1.53  & 67.27   $\pm$    1.78  & 62.24   $\pm$    2.62  & 70.73   $\pm$    0.98  & 65.54 \\
 110  & 60.05   $\pm$    1.27  & 56.60   $\pm$    1.48  & 52.77   $\pm$    2.11  & 59.73   $\pm$    0.81  & 55.28 \\
 115  & 55.42   $\pm$    1.17  & 52.17   $\pm$    1.36  & 48.82   $\pm$    1.92  & 55.16   $\pm$    0.73  & 51.01 \\
 120  & 51.32   $\pm$    1.10  & 48.25   $\pm$    1.24  & 45.32   $\pm$    1.74  & 51.10   $\pm$    0.69  & 47.23 \\
 125  & 47.63   $\pm$    1.00  & 44.73   $\pm$    1.16  & 42.16   $\pm$    1.59  & 47.46   $\pm$    0.62  & 43.83 \\
 130  & 44.33   $\pm$    0.94  & 41.59   $\pm$    1.08  & 39.32   $\pm$    1.45  & 44.19   $\pm$    0.57  & 40.80 \\
 135  & 41.36   $\pm$    0.87  & 38.77   $\pm$    1.00  & 36.77   $\pm$    1.33  & 41.26   $\pm$    0.53  & 38.07 \\
 140  & 38.67   $\pm$    0.81  & 36.22   $\pm$    0.93  & 34.45   $\pm$    1.23  & 38.60   $\pm$    0.49  & 35.60 \\
 145  & 36.23   $\pm$    0.77  & 33.92   $\pm$    0.87  & 32.36   $\pm$    1.13  & 36.21   $\pm$    0.46  & 33.37 \\
 150  & 34.02   $\pm$    0.71  & 31.83   $\pm$    0.81  & 30.46   $\pm$    1.04  & 34.03   $\pm$    0.43  & 31.34 \\
 155  & 32.00   $\pm$    0.67  & 29.93   $\pm$    0.77  & 28.72   $\pm$    0.97  & 32.04   $\pm$    0.40  & 29.49 \\
 160  & 30.16   $\pm$    0.64  & 28.20   $\pm$    0.72  & 27.14   $\pm$    0.90  & 30.22   $\pm$    0.38  & 27.81 \\
 165  & 28.48   $\pm$    0.62  & 26.62   $\pm$    0.68  & 25.70   $\pm$    0.83  & 28.58   $\pm$    0.36  & 26.28 \\
 170  & 26.93   $\pm$    0.57  & 25.16   $\pm$    0.65  & 24.37   $\pm$    0.78  & 27.05   $\pm$    0.34  & 24.87 \\
 175  & 25.52   $\pm$    0.54  & 23.83   $\pm$    0.61  & 23.15   $\pm$    0.73  & 25.65   $\pm$    0.32  & 23.58 \\
 180  & 24.21   $\pm$    0.52  & 22.61   $\pm$    0.58  & 22.03   $\pm$    0.69  & 24.37   $\pm$    0.31  & 22.39 \\
 185  & 23.00   $\pm$    0.49  & 21.48   $\pm$    0.56  & 20.99   $\pm$    0.64  & 23.18   $\pm$    0.29  & 21.30 \\
 190  & 21.90   $\pm$    0.47  & 20.44   $\pm$    0.53  & 20.04   $\pm$    0.61  & 22.09   $\pm$    0.28  & 20.29 \\
 200  & 19.91   $\pm$    0.43  & 18.59   $\pm$    0.49  & 18.33   $\pm$    0.55  & 20.14   $\pm$    0.26  & 18.49 \\
 220  & 16.75   $\pm$    0.37  & 15.64   $\pm$    0.41  & 15.59   $\pm$    0.45  & 17.03   $\pm$    0.22  & 15.62 \\
 240  & 14.38   $\pm$    0.33  & 13.44   $\pm$    0.36  & 13.54   $\pm$    0.38  & 14.70   $\pm$    0.19  & 13.46 \\
 260  & 12.60   $\pm$    0.29  & 11.79   $\pm$    0.33  & 12.01   $\pm$    0.34  & 12.96   $\pm$    0.18  & 11.85 \\
 280  & 11.27   $\pm$    0.27  & 10.56   $\pm$    0.30  & 10.86   $\pm$    0.30  & 11.66   $\pm$    0.17  & 10.64 \\
 300  & 10.32   $\pm$    0.25  &  9.69   $\pm$    0.28  & 10.07   $\pm$    0.29  & 10.75   $\pm$    0.16  &  9.80 \\
\hline
\end{tabular}
\normalsize
\caption{\label{T6}
\small
NNLO predictions for the production cross sections $\sigma(pp\to H^0+X)$ [$pb$] at LHC for $\sqrt{S} = 14$~TeV.  
The errors refer to the $\pm 1 \sigma$ pdf uncertainties.}  
\end{center}
\end{table}
%\vfill
%\end{landscape}
%%%%%%%%%%%%%%%%%%%%%%%%%%%%%%%%

\begin{figure}[H] 
\begin{center}
\epsfig{figure=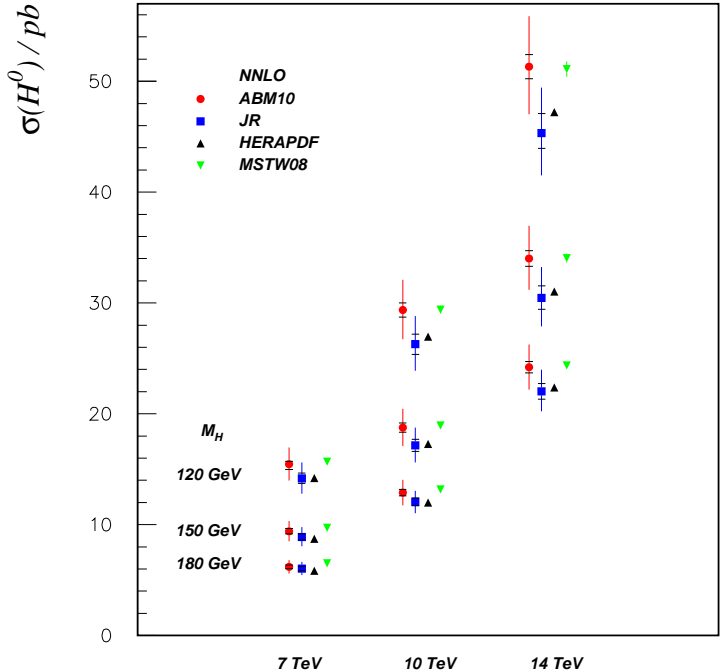,width=0.55\linewidth} 
\end{center}
\caption[]{\label{FIG:HIG} \small Predictions of the 
inclusive Higgs--boson production cross sections at NNLO for different energies at the LHC for the parton 
distributions ABM10, JR, HERAPDF, MSTW08, \cite{ABM,JR,HERAPDF,MSTW08}. For the ABM10 and 
JR distributions the scale variation errors corresponding to the range $M_H/2 \leq \mu_F = \mu_R \leq 2 M_H$ 
are 
included. The inner error bars refer to the pdf--errors only.}
\end{figure}

In Figure~\ref{FIG:HIG} we illustrate the different predictions of ABM10, JR, HERAPDF, and MSTW08 
for the inclusive Higgs production cross section at the LHC for the energies $\sqrt{S} = 7, 10$ and 14~TeV.
The cross sections rise roughly $\propto \sqrt{S}$. In this energy range ABM10 and MSTW08 predict nearly equal 
Higgs boson production cross sections, while those by JR and HERAPDF are found to be lower. For ABM10 and JR we 
include in the error also the scale variation uncertainty varying $\mu_R = \mu_F$ in the range
$[M_H/2, 2M_H]$ and added the pdf--errors in quadrature. It should be kept in mind that the scale variation errors are 
correlated when comparing the predictions for the
different pdf sets. 
Although the various NNLO predictions differ w.r.t. the present pdf--errors, they are consistent within $1 \sigma$ taking the scale 
variation errors
into account.  At NNLO these are still significant, which will require to account for {\it even} higher order 
corrections, despite the well-known fact that soft gluon resummation 
leads to improvements (see \cite{H0NNLO} and references therein).
%----------------------------------------------------------------------------------
%----------------------------------------------------------------------------------

The typical scale uncertainty of each individual NNLO prediction amounts to about $\pm 9 \%$ at 7 TeV and $\pm 
8 \%$ at 14 TeV which {\it doubles} at NLO \cite{JR,HARL}. 
A measure for the uncertainty of the Higgs boson production cross sections is thus obtained
adding in quadrature the largest difference of central values and the largest error, including scale variations, of one of the 
predictions, and dividing this result by the central value of the smallest prediction. Higgs boson production at LHC can be 
predicted with an accuracy of about 
10--17 \% at NNLO~\footnote{Notice that the largest uncertainty refers to the predictions for $M_H = 120$ GeV at 14 TeV, and the 
uncertainty decreases for $M_H$ increasing; the smallest uncertainty of about 10 \% refers to the predictions for $M_H = 180$ 
GeV at 7 TeV.} (with a total uncertainty being almost {\it twice} as large at NLO), whereas 
the uncertainty almost doubles at the Tevatron ($\sqrt{S} = 1.96$ TeV). Furthermore, the NNLO predictions are typically about 
20~\% larger than at NLO.

We finally remark that a possible way to determine experimentally the gluon density in the mass region 
relevant 
for Higgs boson production consists in a precise measurement of the single top quark production cross section 
\cite{STOP1,POWEG,STOP2}  for 
scales around $\mu \sim m_t$. At present somewhat differing results for this process were obtained at 
Tevatron 
\cite{STOP_exp}, however, still with large errors,
%----------------------------------------------------------------------------------
{\small
\begin{eqnarray} 
\sigma(p\overline{p} \rightarrow t(\bar{t})+\bar{b}(b) + X) &=& 2.3 \begin{array}{c} + 
\text{\fs 0.6} \\ \text{\fs -- 0.5} \end{array} \text{(stat. + sys.)}~~pb~~\text{CDF} 
\\
%\end{eqnarray} 
%\begin{eqnarray} 
&=& 4.19 
\begin{array}{c} + \text{\fs 1.24} \\ 
\text{\fs -- 1.14} \end{array} \text{(stat. + sys.)}~~pb~~\text{D0} 
\end{eqnarray}} 
%---------------------------------------------------------------------------------- 
with~\footnote{Assuming unitarity for the CKM matrix would imply a far smaller error 
\cite{PDG2010}.}~~$V_{tb} 
= 0.91 \pm 0.11$.
%---------------------------------------------------------------------------------- 
Theoretical NLO QCD predictions~\footnote{We would like to thank S. Alioli for computing the reference values 
for single top production
with {\tt POWEG} \cite{POWEG}.} are
%---------------------------------------------------------------------------------- 
\begin{eqnarray} 
\label{st1}
\sigma(p\overline{p} \rightarrow t(\bar{t})+\bar{b}(b) + X) &=& |V_{tb}|^2 (2.420 \pm 
0.003)~pb~~~\text{MSTW08}
\\
\label{st2}
                                                           &=&  |V_{tb}|^2 (2.195 \pm 
0.003)~pb~~~\text{ABKM09}~,
\end{eqnarray}
which differ by about 10\%. The NLO scale variation errors according to $m_t/2 < \mu_{F,R} < 2 m_t$ amount up 
to  $10\%$ at Tevatron, are lower than 5 \% at LHC \cite{STOP2}, and can be further improved at NNLO.
The different values in (\ref{st1}) and (\ref{st2}) are caused partly by yet different gluon 
distributions in the region $x \sim 0.1$, see Figure~2, but at least to the same extent they are due 
to the different values of $\alpha_s$, cf.~Table~1. Precision measurements of the single top production cross 
sections at the LHC, although challenging, with an accuracy of $5\%$ or better would allow to determine the 
gluon density in the same experiments which measure the production of the Higgs boson as well. 

\vspace*{5mm}
\noindent
{\large \sf {4~~Conclusions.}}~~
The present analysis shows that the current NNLO analyses of the world deep-inelastic data, supplemented
with other relevant hard scattering data to measure the parton distributions, yield still different 
results at the level of less than about 10 \% for $W^\pm$ and $Z^0$ boson production, and at the level of 
$10-17 \%$ for Higgs boson production at the LHC. (These uncertainties would almost {\it double} at NLO.)
The  present variations in the predictions based on \cite{ABM,ABKM,JR,HERAPDF,MSTW08,MSTW10}
are both due to the QCD-scale $\Lambda_{\rm QCD}$ (or $\alpha_s(M_Z^2)$) and differences in the parton densties, 
as well as due to the renormalization and factorization scale uncertainties of the theoretical predictions. 
Future detailed work is needed to study and to further delineate these differences. The 
differences documented in this study form essential contributions to the theory error of the measurements of these scattering cross 
sections at the Tevatron and the LHC, which should therefore be compared to all the predictions based on the 
distributions ABM10 \cite{ABM}, JR \cite{JR}, MSTW08 \cite{MSTW08}, and HERAPDF \cite{HERAPDF} at NNLO.

\vspace*{4mm}
\noindent
{\bf Acknowledgment.}~
We would like to thank ECT* at Trento for hospitality and providing an inspiring environment at
Villa Tambosi. 
For discussions we would like to thank S. Alioli, H.~B\"ottcher, M.~Cooper-Sarkar, G.~Dissertori, A.~Glazov, 
and M.~Grazzini. This work was supported in part by DFG Sonderforschungsbereich Transregio 9, 
Computergest\"utzte Theoretische Teilchenphysik, Helmholtz Alliance, the European Commission MRTN HEPTOOLS under 
the Contract No. MRTN-CT-2006-035505, and the Swiss National Science Foundation (SNF) under the Contract 200020-126691.

%%%%%%%%%%%%%%%%%%%%%%%%%%%%%%%%%%%%%%%%%%%%%%%%%%%%%%%%%%%%%%%%%%%%%%%%%%%%%%%%%%%%%%%%%%%%%%%%
%\renewcommand{\theequation}{\ref{sec:appA}.\arabic{equation}}
\renewcommand{\theequation}{A.\arabic{equation}}
\setcounter{equation}{0}
%%%%%%%%%%%%%%%%%%%%%%%%%%%%%%%%%%%%%%%%%%%%%%%%%%%%%%%%%%%%%%%%%%%%%%%%%%%%%%%%%%%%%%%%%%%%%%%%
\appendix

\vspace*{5mm}
\noindent
{\large \sf {Appendix.}}~~
%\section{Summary of the experimental results}
%%%%%%%%%%%%%%%%%%%%%%%%%%%%%%%%%%%%%%%%%%%%%%%%%%%%%%%%%%%%%%%%%%%%%%%%%%%%%%%%%%%%%%%%%%%%%%%%
%
%\vspace{1mm}
%\noindent
In the following we summarize the different experimental measurements of the $W^\pm$ and $Z^0$-boson
production cross sections in $p\overline{p}$ and $pp$ collisions for 
comparison with the predictions based on the current NNLO parton densities.

The cross sections for $p\overline{p}$ collisions measured by UA1, UA2 and D0 experiments are (averaging over the 
different channels 
measured by UA1 for $Z^0$ production)~:
%----------------------------------------------------------------------------------
\begin{eqnarray}
\text{UA1(0.546 TeV):}~~~
\sigma^{W^+ + W^-} B(W^\pm \rightarrow \ell \nu) &=&
0.55 \pm 0.12~~~\text{$nb$}~~~          \\  
\sigma^{Z^0}B(Z^0 \rightarrow \ell^+\ell^-) &=&
0.070 \pm 0.046~~~\text{$nb$}~~~ \\
%----            
\text{UA2(0.546 TeV):}~~~  
\sigma^{W^+ + W^-} B(W^\pm \rightarrow \ell \nu) &=&
0.57 \pm 0.08~~~\text{$nb$}~~~       \\  
\sigma^{Z^0}B(Z^0 \rightarrow \ell^+\ell^-) &=&
0.116 \pm 0.041~~~\text{$nb$}~~~ 
\\
\text{UA1(0.630 TeV):}~~~ 
\sigma^{W^+ + W^-} B(W^\pm \rightarrow \ell \nu) &=&
0.63 \pm 0.13~~~\text{$nb$}~~~         \\  
\sigma^{Z^0} {B}(Z^0 \rightarrow \ell^+\ell^-) &=& 
0.070 \pm 0.019~~~\text{$nb$}~~~ 
%----            
\\
%\end{eqnarray}
%\begin{eqnarray}
%----            
\text{UA2(0.630 TeV):}~~~ 
\sigma^{W^+ + W^-} B(W^\pm \rightarrow \ell \nu) &=&
0.57 \pm 0.08~~~\text{$nb$}~~~         \\  
\sigma^{Z^0} {B}(Z^0 \rightarrow \ell^+\ell^-) &=& 
0.073 \pm 0.016~~~\text{$nb$}~~~ \\ 
%----            
\text{D0(0.630 TeV):}~~~ 
\sigma^{W^+ + W^-} B(W^\pm \rightarrow \ell \nu) &=&
0.658 \pm 0.058 \pm 0.034~~~\text{$nb$}~.~~~          
\end{eqnarray}
%----------------------------------------------------------------------------------
The cross section measurement of CDF at $\sqrt{S} = 0.630$ TeV has not been 
published~\cite{EJ}. 

At similar energies of $\sqrt{S} = 0.5$~TeV the inclusive $W^\pm$ boson cross sections 
were recently measured by the PHENIX experiment \cite{PHENIX} in $pp$ collisions at RHIC~:
\begin{flushleft}
\text{PHENIX(0.5 TeV):}~~~ 
\end{flushleft} 

\vspace*{-3mm}
%----------------------------------------------------------------------------------
\begin{eqnarray}
\hspace*{-3mm}
\sigma^{W^+} {B}(W^+ \rightarrow e^+ \nu) &=& 
0.1441\pm 0.0212  \begin{array}{c} +0.0034\\-0.0103 \end{array} \pm 15\% (\text{norm})~~~\text{$nb$}~~~          
\\  
%----            
\hspace*{-3mm}
\sigma^{W^-} {B}(W^- \rightarrow e^- \overline{\nu}) &=& 
0.0317\pm 0.0212  \begin{array}{c} +0.0101\\-0.0082 \end{array} \pm 15\% (\text{norm})~~~\text{$nb$}~.~~          
\end{eqnarray}
%----------------------------------------------------------------------------------
Here and in the following the errors are given in the sequence of statistical, systematic, 
and luminosity/normalization errors. 

The $p\overline{p}$ experiments CDF and D0 at the Tevatron performed inclusive measurements for $W^\pm$ and 
$Z^0$
production at $\sqrt{S} = 1.8$ and 1.96 TeV~:
\begin{flushleft}
\text{CDF(1.8 TeV)} \text{\cite{CDF} :} 
\end{flushleft} 

\vspace*{-3mm}
%----------------------------------------------------------------------------------
\begin{eqnarray}
\label{eq:TEV1}
\sigma^{W^+ + W^-} B(W^\pm \rightarrow \ell \nu) &=&
2.49 \pm 0.12~~~\text{$nb$}~~~            \\  
%------
\sigma^{Z^0}B(Z^0 \rightarrow \ell^+\ell^-) &=&
0.231 \pm 0.012~~~\text{$nb$}~~~ 
\end{eqnarray}
%----------------------------------------------------------------------------------
\begin{flushleft}
\text{D0(1.8 TeV)} \text{\cite{D0} :}~~~ 
\end{flushleft} 

\vspace*{-3mm}
%----------------------------------------------------------------------------------
\begin{eqnarray}
\sigma^{W^+ + W^-} B(W^\pm \rightarrow \ell \nu) &=&
2.310 \pm 0.001 \pm 0.005 \pm 0.100~~~\text{$nb$}~~~            \\  
%------
\sigma^{Z^0}B(Z^0 \rightarrow \ell^+\ell^-) &=&
0.221 \pm 0.003 \pm 0.004 \pm 0.010~~~\text{$nb$}~~~                      
\end{eqnarray}
%----------------------------------------------------------------------------------
\begin{flushleft}
\text{CDF(1.96 TeV)} \text{\cite{CDF,CDF1} :}~~~ 
\end{flushleft} 

\vspace*{-3mm}
%----------------------------------------------------------------------------------
\begin{eqnarray}
\sigma^{W^+ + W^-} B(W^\pm \rightarrow \ell \nu) &=&
%------
2.749 \pm 0.010  \pm 0.053 \pm 0.165~~~\text{$nb$}~~~           \\  
\sigma^{Z^0}B(Z^0 \rightarrow \ell^+\ell^-) &=&
%------
0.2549 \pm 0.0033 \pm 0.0045 \pm 0.0152~~~\text{$nb$}~~~
\\                       
\label{eq:TEV2}
&=& 0.2566 \pm 0.0007 \pm 0.0020 \pm 0.0154~~~\text{$nb$}.~~~                       
\end{eqnarray}
%----------------------------------------------------------------------------------

First measurements of the inclusive $W^\pm$ and $Z^0$ boson cross sections at LHC were performed
at $\sqrt{S}$ = 7 TeV~:
%----------------------------------------------------------------------------------
\begin{eqnarray}
%\text{CMS:} \text{\cite{CMS}}~~~
%\sigma^{W^+ + W^-} B(W^\pm \rightarrow \ell \nu) &=&
%9.22 \pm 0.24 \pm 0.47 \pm 1.01~~~\text{$nb$}         \\  
%\sigma^{W^+ } {B}(W \rightarrow \ell \nu) &=& 
%         5.50 \pm 0.18 \pm 0.29 \pm 0.61~~~\text{$nb$}  \\ 
%\sigma^{W^-} {B}(W \rightarrow \ell \nu) &=& 
%         3.60 \pm 0.13 \pm 0.19 \pm 0.40~~~\text{$nb$}\\                       
%\sigma^{Z^0}B(Z^0 \rightarrow \ell^+\ell^-) &=&
%0.882 
%\begin{array}{c} + 0.077\\ -0.073 \end{array}~ 
%\begin{array}{c} + 0.042\\ -0.036 \end{array}~ 
%\pm 0.097~~~\text{$nb$} \\                       
\text{CMS} \text{\cite{CMS}}:~~~
\sigma^{W^+ + W^-} B(W^\pm \rightarrow \ell \nu) &=&
9.951 \pm 0.073 \pm 0.280 \pm 1.095~~~\text{$nb$}~~~         \\  
\sigma^{W^+ } {B}(W^+ \rightarrow \ell^+ \nu) &=& 
         5.859\pm 0.059 \pm 0.168 \pm 0.645~~~\text{$nb$}~~~  \\ 
\sigma^{W^-} {B}(W^- \rightarrow \ell^- \overline{\nu}) &=& 
         4.140 \pm 0.064 \pm 0.254 \pm 0.455~~~\text{$nb$}~~~\\                       
\sigma^{Z^0}B(Z^0 \rightarrow \ell^+\ell^-) &=&
0.960 \pm 0.037 \pm 0.059 \pm 0.106
%\begin{array}{c} + 0.077\\ -0.073 \end{array}~ 
%\begin{array}{c} + 0.042\\ -0.036 \end{array}~ 
%\pm 0.097
~~~\text{$nb$}~~~ \\                       
\text{ATLAS} \text{\cite{ATLAS}}:~~~
\sigma^{W^+ + W^-} B(W^\pm \rightarrow \ell \nu) &=&
9.96 \pm 0.23 \pm 0.50 \pm 1.10~~~\text{$nb$}~~~         \\  
\sigma^{W^+ } {B}(W^+ \rightarrow \ell \nu) &=& 
         5.93 \pm 0.17 \pm 0.30 \pm 0.65~~~\text{$nb$}~~~  \\ 
\sigma^{W^-} {B}(W^- \rightarrow \ell^- \overline{\nu}) &=& 
          4.00  \pm 0.15 \pm 0.20 \pm 0.44~~~\text{$nb$}~~~  \\                       
\sigma^{Z^0}B(Z^0 \rightarrow \ell^+\ell^-) &=&
0.82 \pm 0.06 \pm 0.05 \pm 0.09~~~\text{$nb$}.~~~                      
\end{eqnarray}
%----------------------------------------------------------------------------------
These cross sections correspond to analyzed samples of $2.88~{\it pb}^{-1}$ and $ 0.32~{\it pb}^{-1}$, while 
the current run may accumulate $O(50~{\it pb}^{-1})$, which will lead to a significant 
reduction of the statistical error and allow to improve the systematic errors.

To compare with the theoretical predictions one may use the current measured branching 
fractions~\cite{PDG2010}
%----------------------------------------------------------------------------------
\begin{alignat}{8}
&{B}(W^+ \rightarrow e^+ \nu_e)       ~~&=&~~ 0.1075  \pm 0.0013~,~~~~   
&{B}(W^+ \rightarrow \mu^+ \nu_\mu)   ~~&=&~~ 0.1057  \pm 0.0015~,~~~~\nonumber\\   
&{B}(W^+ \rightarrow \tau^+ \nu_\tau) ~~&=&~~ 0.1125  \pm 0.0020~, & & &  
\end{alignat}
with 
$B(W^- \rightarrow \ell^- \overline{\nu}_l)  
= 
B(W^+ \rightarrow \ell^+ \nu_l)$ 
and
\begin{alignat}{8}
&{B}(Z^0 \rightarrow e^+ e^-)        ~~&=&~~ 0.03363 \pm 0.00004~,~~~ 
&{B}(Z^0 \rightarrow \mu^+ \mu^-)    ~~&=&~~ 0.03366 \pm 0.00007~, & & & \nonumber\\ 
&{B}(Z^0 \rightarrow \tau^+ \tau^-)  ~~&=&~~ 0.03367 \pm 0.00008~, 
\end{alignat}
%----------------------------------------------------------------------------------
or their corresponding averages.

\vspace*{-5mm}
%\newpage
%%%%%%%%%%%%%%%%%%%%%%%%%%%%%%%%%%%%%%%%%%%%%%%%%%%%%%%%%%%%%%%%%%%%%%%

\end{document}